\documentclass[12pt]{article}

\usepackage{scicite}
\usepackage{amsmath}
\usepackage{times}
\usepackage{url}

\newenvironment{sciabstract}{%
\begin{quote} \bf}
{\end{quote}}

\title{Entangled Probability Distributions for Center-of-Mass
Tomography\footnote{We dedicate this paper to the memory of Professor
Francesco Saverio Persico, famous scientist and respectable noble person, with
whom Margarita Man'ko and Vladimir Man'ko had non-forgettable meetings and
discussions about quantum physics during their visits to the University of
Palermo.}}

\author{Ivan V. Dudinets $^{1}$*\orcidA{}, Margarita A. Man'ko $^{2}$ and Vladimir I. Man'ko $^{2}$}

\author
{Ivan V. Dudinets,$^{1\ast}$ Margarita A. Man'ko $^{2}$ and Vladimir I. Man'ko $^{2}$\\
\\
\normalsize{$^{1}$ Russian Quantum Center, Skolkovo, Moscow 121205, Russia;}\\
\normalsize{$^{2}$ Lebedev Physical Institute, Russian Academy of Sciences,
Leninskii Prospect 53,} \\\normalsize{Moscow 119991, Russia}\\
\\
\normalsize{$^\ast$Correspondence: dudinets@phystech.edu}
}

\date{}

\begin{document} 


\baselineskip24pt


\maketitle


\begin{sciabstract}
  We review the formalism of center-of-mass tomograms that allows us to describe
quantum states in terms of probability distribution functions. We introduce
the concept of separable and entangled probability distributions for the
center-of-mass tomography. We obtain the time evolution of center-of-mass
tomograms of entangled states of the inverted oscillator.
\end{sciabstract}


\section*{Introduction}


The conventional probability theory~\cite{kolmogoroff1933grundbegriffe}
provides the basis to study the properties of quantum systems. The formulation
of quantum mechanics is based on the Schr{\"o}dinger
equation~\cite{schrodinger1926quantisierung}. The states of quantum systems
are usually described by the complex wave function or the density
operator~\cite{landau1927dampfungsproblem,neumann1927wahrscheinlichkeitstheoretischer,landau2013quantum,dirac1981principles,schrodinger1926quantisierung}.
An alternative way to represent the states is given by quasi-probability
distributions like the Wigner
function~\cite{wigner1932quantum,schleich2011quantum}, the Husimi
$Q$-function~\cite{husimi1940some,kano1965new}, and the Glauber--Sudarshan
$P$-function~\cite{glauber1963coherent,sudarshan1963equivalence}. There exist
other quasi-probability
functions~\cite{kirkwood1933quantum,margenau1961correlation,cohen1966generalized,cohen1989time}.
The quasi-probability distributions are not fair probabilities; they are
functions on the phase space and their arguments, position and momentum, are
not simultaneously measurable due to the uncertainty
relations~\cite{heisenberg1927anschaulichen,robertson1930general,
schrodinger1930sitzungsberichte}. In addition, the quasi-probability functions
can take negative or even complex values~\cite{schleich2011quantum}.  The
probability representation of quantum
mechanics~\cite{mancini1996symplectic,ibort2009introduction} 
provides the description of the states of quantum systems in terms of
nonnegative probability distributions called tomograms, both for discrete
variables~\cite{dodonov1997positive}
and continuous variables~\cite{mancini1996symplectic,ibort2009introduction}.
The tomograms are related to the density operator or quasi-probability
distributions by means of invertible integral transforms. For example, the
Wigner function and tomogram are connected by the Radon
transform~\cite{ibort2009introduction}. All invertible maps of density
operators and observables onto functions are described by the star-product
formalism~\cite{man2002alternative}. Special cases of such maps are the
sympletic tomography~\cite{mancini1996symplectic,ibort2009introduction}
and the center-of-mass tomography
~\cite{arkhipov2003tomography,arkhipov2005quantum}.
Examples of the center-of-mass
tomograms~\cite{arkhipov2003tomography,dudinets2018center} and symplectic
tomograms for states of a harmonic oscillator like Fock states and coherent or
Schr{\"o}dinger cat states are presented
in~\cite{ibort2009introduction,man2023quantum}. The tomographic methods and
their applications to analyze quantum systems were considered
in~\cite{d2003quantum,filinov2008center,lvovsky2009continuous,bazrafkan2009stationary,toninelli2019concepts}.
Some other aspects of quantum systems in the context of interactions with
light and external
fields~\cite{compagno1995atom,carbonaro1979canonical,benivegna1994new,cirone2007casimir}
and the entanglement
formation~\cite{migliore2006generation,grimaudo2022greenberger} were
discussed.

The tomographic picture allows one to describe the states of both quantum and
classical systems by
tomograms~\cite{mancini1996symplectic,ibort2009introduction}; it finds
application in cosmology~\cite{stornaiolo2020emergent,berra2022tomography}.
The difference between quantum systems and classical systems consists in the
possibility to be in an entangled state. This fact leads to the concept of
entangled probability distributions~\cite{chernega2023dynamics} that has not
been studied in classical probability theory. The classical probability theory
is described, for example,
in~\cite{holevo2011probabilistic,kolmogorov2018foundations}. Some new aspects
of entanglement phenomena are discussed in recent
papers~\cite{zanardi2001virtual, zanardi2004quantum, basieva2022conditional, khrennikov2023entanglement,khrennikov2019classical,khrennikov2021roots}. In this paper, we introduce the notion of  entangled probability distributions for the center-of-mass tomography and study them in view of the probability representation of quantum mechanics. We consider examples of the center-of-mass tomograms of entangled states of harmonic and inverted oscillators. We also determine the time evolution of these states using the method of integrals of motion developed in~\cite{man2023quantum}. 
The idea of the method is that, for systems like harmonic oscillators or
inverted oscillators, the position and momentum operators in the Heisenberg
representation are expressed in terms of integrals of motion and linear in the
position and momentum operators. This allows one to obtain the evolution of
tomograms by corresponding time-dependent transforms of the parameters of
initial tomograms.  Also, we consider examples of the cluster
tomograms~\cite{dudinets2018center} -- the generalization of center-of-mass
tomography.

The aim of our work is to construct new conditional probability distribution
functions and entangled probability distribution functions, which describe
quantum states of quantum systems, and to study their properties. Earlier,
these probability distributions were not considered, since classical and
quantum systems provide different randomness phenomena such as, for example,
the existence of the position-and-momentum uncertainty relations for quantum
systems~\cite{heisenberg1927anschaulichen, robertson1930general,
schrodinger1930sitzungsberichte}.

This paper is organized as follows.

In Sec.~2, we review the probability representation of quantum states paying a
special attention to the center-of-mass tomography. In Sec.~3, we consider
examples of the center-of-mass tomograms for entangled states and its
connection to symplectic tomograms.  In Sec.~4, we obtain the dynamics of
tomograms for harmonic and inverted oscillators. Section~5 is devoted to the
cluster tomography. Summary and prospects are given in Sec.~6.
\section{Entangled Probability Distributions}

We consider a quantum system with two degrees of freedom. In the
center-of-mass tomography, the state of a quantum system is described by the
center-of-mass tomogram. The center-of-mass tomogram of a quantum state with the
density operator $\hat{\rho}$ is defined as
follows~\cite{arkhipov2003tomography}:
\begin{equation}
    w(X|\mu_1,\nu_1,\mu_2,\mu_2) = \mbox{Tr} \left(\hat{\rho}\,\delta
    \left(X-\mu_1 \hat{q}_1-\nu_1 \hat{p}_1-\mu_2 \hat{q}_2-\nu_2 \hat{p}_2\right)\right),
    \label{eq:tomogram}
\end{equation}
where $\hat{q}_j$ and $\hat{p}_j$ are the position and momentum operators for
each degree of freedom. The center-of-mass tomogram is a nonnegative
probability distribution of the random variable $X$ associated with the
center-of-mass position of the system in the phase space in rotated and scaled
reference frames, which are determined by parameters $\mu_1$, $\nu_1$,
$\mu_2$, and $\nu_2$.  Note that the tomogram effectively operates with four
variables, due to the homogeneity property of the Dirac delta-function. The
density operator of a state can be reconstructed from the center-of-mass
tomogram; it reads
\begin{eqnarray}
    \hat{\rho} = \frac{1}{4\pi^2}\int w(X|\mu_1,\nu_1,\mu_2,\mu_2)
    \exp\left({i(X-\mu_1 \hat{q}_1-\nu_1 \hat{p}_1-\mu_2 \hat{q}_2-\nu_2 \hat{p}_2)}\right)
    \nonumber\\
    \times\,dX
    \,d\mu_1\,d\nu_1\,d\mu_2\,d\nu_2.
    \label{eq:inverted}
\end{eqnarray}

The center-of-mass tomogram can be treated as the conditional probability
distribution~\cite{dudinets2018center}, where $\mu_1$, $\nu_1$, $\mu_2$, and
$\nu_2$ are parameters describing the condition of measuring $X$. The
treatment follows from the ``no-signalling'' property~\cite{beautyInphys},
\begin{equation}
    \int\,dX\,w(X|\mu_1,\nu_1,\mu_2,\nu_2)=1,
\end{equation}
that holds for any parameters  $\mu_1$, $\nu_1$, $\mu_2$, and $\nu_2$. In the
case of pure states, the center-of-mass tomogram is given in
\cite{arkhipov2003tomography}; it is
\begin{eqnarray}
    w(X|\mu_1,\mu_2,\nu_1,\nu_2) =
    \int dY_1dY_2\frac{\delta(X-Y_1-Y_2)}{4\pi^2|\nu_1\nu_2|}\nonumber\\
    \times\left|\int dq_1 dq_2\,\psi(q_1,q_2)\exp
    \left(
    \frac{i\mu_1}{2\nu_1}q^2_1+\frac{i\mu_2}{2\nu_2}q^2_2-\frac{iY_1}{\nu_1}q_1-\frac{iY_2}{\nu_2}q_2
    \right)
    \right|^2.
    \label{eq:pure}
\end{eqnarray}

There exist other probability distributions that can be identified with
quantum states; for instance, symplectic tomogram. Symplectic tomogram is the
nonnegative probability distribution of random variables $X_1$ and $X_2$
associated with the position of the system in the phase space in rotated and
scaled reference frames determined by parameters $\mu_1$, $\nu_1$, $\mu_2$,
and $\nu_2$; it reads
\begin{equation}
     w^{s}(X_1,X_2|\mu_1,\mu_2,\nu_1,\nu_2)=\mbox{Tr} \left(\hat{\rho}\,\delta\left(X_1-\mu_1 \hat{q}_1-
     \nu_1 \hat{p}_1\right)\,\delta\left(X_2-\mu_2 \hat{q}_2-\nu_2 \hat{p}_2\right)\right).
    \label{eq:symplectic}
\end{equation}
The inverse transform is
\begin{eqnarray}
    \hat{\rho} = \frac{1}{4\pi^2}\int w^{s}(X_1,X_2|\mu_1,\mu_2,\nu_1,\nu_2)\,
    \exp\left({i(X_1+X_2-\mu_1 \hat{q}_1-\nu_1 \hat{p}_1-\mu_2 \hat{q}_2-\nu_2 \hat{p}_2)}\right)
    \nonumber\\\times
    \,dX_1\,dX_2
    \,d\mu_1\,d\nu_1\,d\mu_2\,d\nu_2.
\end{eqnarray}

The center-of-mass tomogram and symplectic tomogram are related as follows:
\begin{equation}
    w^{s}(X_1,X_2|\mu_1,\mu_2,\nu_1,\nu_2) = \frac{1}{4\pi^2}\int
    w(X|k_1\mu_1,k_2\mu_2,k_1\nu_1,k_2\nu_2)e^{i(X-k_1X_1-k_2X_2)}\,
    dk_1\,dk_2\,dX.
\end{equation}
The state of the first subsystem can be found in terms of the center-of-mass
tomogram of the whole system, in view of the formula~\cite{dudinets2018center}
\begin{equation}
    w_1(X_1|\mu_1,\nu_1) = \mbox{Tr} \left(\hat{\rho}_1\,\delta\left(X_1-\mu_1 \hat{q}_1-\nu_1 \hat{p}_1\right)\right)=
    \frac{1}{2\pi}\int w(X|k\mu_1,0,k\nu_1,0)e^{i(X-kX_1)}\,dk\,dX,
\end{equation}
where $\hat{\rho}_1 = \mbox{Tr}_2\,\hat{\rho}$ is the density operator of the first subsystem obtained by taking the partial trace of the density operator $\hat{\rho}$ of the whole system over the second subsystem.

Let us introduce the concept of separable and entangled
probability distributions for the center-of-mass tomography. 
The symplectic tomogram of a separable state of a system, which consists of
two subsystems, is represented by the convex sum of symplectic tomograms of
subsystems~\cite{chernega2023dynamics},
\begin{equation}
    w^{s}(X_1,X_2|\mu_1,\mu_2,\nu_1,\nu_2) = \sum_k p_k \, w^{(k)}_{1}(X_1|\mu_1,\nu_1)\, w^{(k)}_2(X_2|\mu_2,\nu_2),
    \label{eq:s_separable}
\end{equation}
where $p_k$ are probabilities, i.e., $p_k\geq 0$ and $\sum_k p_k = 1$. The probability distribution $ w^{s}(X_1,X_2|\mu_1,\mu_2,\nu_1,\nu_2)$ is called the entangled probability distribution if it cannot be presented as the convex sum of the form~(\ref{eq:s_separable})~\cite{chernega2023dynamics}. For the center-of-mass 
tomogram, the relation between tomograms of the system and its subsystems for a separable state follows from (\ref{eq:s_separable}); it 
has the form
\begin{equation}
    w(X|\mu_1,\mu_2,\nu_1,\nu_2) = \sum_k p_k \,\int  w^{(k)}_1(X_1|\mu_1,\nu_1)\, w^{(k)}_2(X-X_1|\mu_2,\nu_2)\,dX_1.
    \label{eq:cm_separable}
\end{equation}
The probability distribution $ w(X|\mu_1,\mu_2,\nu_1,\nu_2)$ is said to be the entangled probability distribution if it cannot be cast in the form~(\ref{eq:cm_separable}). The generalization of formula~(\ref{eq:cm_separable}) to the case of systems with many degrees of freedom is given in Appendix A. 

In the next section, we consider examples of separable and entangled 
probability distributions for the center-of-mass tomography.

\section{Examples of Entangled Probability Distribution}

Let us consider an entangled state of a two-dimensional oscillator, which is a
superposition of the ground state $\psi_0(q) = \pi^{-1/4}e^{-q^2/2}$ and the
first excited state $\psi_1(q)=\pi^{-1/4}\sqrt{2}\,q\,e^{-q^2/2}$ of the form
\begin{equation}
    \psi_{\text{ent}}(q_1,q_2) = \frac{1}{\sqrt{2}}\left(\psi_0(q_1)\psi_1(q_2)+
    \psi_1(q_1)\psi_0(q_2)\right) = \frac{q_1+q_2}{\sqrt{\pi}}\exp\left(-\frac{q^2_1}{2}-\frac{q^2_2}{2}\right).
    \label{eq:psi_ent}
\end{equation}
The center-of-mass tomogram of this state follows from the general
relation~(\ref{eq:pure}); it reads
\begin{eqnarray}
    w_{\text{ent}}(X|\mu_1,\mu_2,\nu_1,\nu_2) =
\frac{e^{-X^2/\sigma}}{\sqrt{\pi\sigma}}
    \left(\frac{1}{2} -\frac{\mu_1\mu_2+\nu_1\nu_2}{\sigma}+\frac{X^2}{\sigma}
    \left(1+\frac{2(\mu_1\mu_2+\nu_1\nu_2)}{\sigma}\right)
    \right),
\label{eq:cm_ent}
\end{eqnarray}
where $\sigma = \mu_1^2+\mu_2^2+\nu_1^2+\nu_2^2$. We call this probability
distribution the entangled probability distribution, since it determines the
entangled state. To compare it with the center-of-mass tomogram of a separable
state, we consider the following wave function:
\begin{equation}
    \psi_{\text{sep}}(q_1,q_2) = \psi_0(q_1)\psi_1(q_2) = \frac{\sqrt{2}q_2}
    {\sqrt{\pi}}\exp\left(-\frac{q^2_1}{2}-\frac{q^2_2}{2}\right).
    \label{eq:psi_sep}
\end{equation}
The corresponding tomogram is given by
\begin{eqnarray}
    w_{\text{sep}}(X|\mu_1,\mu_2,\nu_1,\nu_2) =
\frac{e^{-X^2/\sigma}}{\sqrt{\pi\sigma^3}}
    \left(
    \mu^2_1+\nu^2_1+\frac{2X^2}{\sigma}\left(\mu^2_2+\nu^2_2\right)
    \right).
\end{eqnarray}

Tomograms of the separable and entangled states of the first susbsystem
considered above are
\begin{equation}
    w^{\text{sep}}_1(X_1|\mu_1,\nu_1) = \frac{1}{\sqrt{\pi(\mu^2_1+\nu^2_1)}}
    \exp{\left(-\frac{X^2_1}{\mu^2_1+\nu^2_1}\right)},
\end{equation}
\begin{equation}
    w^{\text{ent}}_1(X_1|\mu_1,\nu_1) = \frac{1}{\sqrt{\pi(\mu^2_1+\nu^2_1)}}
    \left(\frac{1}{2}+\frac{X^2_1}{\mu^2_1+\nu^2_1}\right)
    \exp{\left(-\frac{X^2_1}{\mu^2_1+\nu^2_1}\right)}.
\end{equation}
Symplectic tomograms of the separable and entangled states read
\begin{equation}
    w_{\text{sep}}^{s}(X_1,X_2|\mu_1,\mu_2,\nu_1,\nu_2) =
    \frac{2X^2_2}{\pi(\mu^2_1+\nu^2_1)^{1/2}(\mu^2_2+\nu^2_2)^{3/2}}
    \exp{\left(-\frac{X^2_1}{\mu^2_1+\nu^2_1}-\frac{X^2_2}{\mu^2_2+\nu^2_2}\right)}.
\end{equation}
\begin{eqnarray}
    w_{\text{ent}}^{s}(X_1,X_2|\mu_1,\mu_2,\nu_1,\nu_2) =
    \frac{1}{\pi(\mu^2_1+\nu^2_1)^{1/2}(\mu^2_2+\nu^2_2)^{1/2}}
    \nonumber\\
    \times
    \left(\frac{X^2_1}{\mu^2_1+\nu^2_1}
    +\frac{X^2_2}{\mu^2_2+\nu^2_2}+
\frac{2X_1X_2\left(\mu_1\mu_2+\nu_1\nu_2\right)}{\left(\mu^2_1+\nu^2_1\right)\left(\mu^2_2+\nu^2_2\right)}
    \right)
     \exp{\left(-\frac{X^2_1}{\mu^2_1+\nu^2_1}-\frac{X^2_2}{\mu^2_2+\nu^2_2}\right)}.
\end{eqnarray}

\section{Dynamics of Tomograms for Hamiltonians Quadratic in the Position and Momentum Operators}
In this section, we consider the evolution of the center-of-mass tomogram of
the systems with Hamiltonians, which are quadratic in the position and
momentum operators. The integrals of motion of such systems are linear in the
position and momentum operators~\cite{markov1987invariants}. This allows one
to obtain the time dependence of the center-of-mass tomogram describing the
quantum state. Indeed, the density operator evolves as
\begin{equation}
    \hat{\rho}(t)=\hat{u}(t)\hat{\rho}(0) \hat{u}^{\dagger}(t),
\end{equation}
where  $\hat{u}(t)=\exp{\left(-it\hat{H}\right)}$ is the evolution operator.
The center-of-mass tomogram corresponding to the state $\hat{\rho}(t)$ is
expressed in terms of the position and momentum operators,  $\hat{q}^H(t) =
\hat{u}^{\dagger}(t)\,\hat{q}\,\hat{u}(t)$ and $\hat{p}^H(t) =
\hat{u}^{\dagger}(t)\,\hat{p}\,\hat{u}(t)$; in the Heisenberg representation,
it is
\begin{eqnarray}
    w(X|\mu_1,\mu_2,\nu_1,\nu_2;t)=\mbox{Tr} \left(\hat{\rho}(t)\,\delta\left(X-\mu_1 \hat{q}_1-\nu_1 \hat{p}_1
    -\mu_2 \hat{q}_2-\nu_2 \hat{p}_2\right) \right)=\nonumber\\
    \mbox{Tr} \left(\hat{\rho}(0)\,\delta\left(X-\mu_1 \hat{q}^H_1(t)-\nu_1 \hat{p}^H_1(t)
    -\mu_2 \hat{q}^H_2(t)-\nu_2 \hat{p}^H_2(t)\right) \right).
\end{eqnarray}
As the first example, let us consider a two-dimensional harmonic oscillator
\begin{equation}
    \hat{H} = \frac{\hat{p}^2_1}{2}+\frac{\hat{q}^2_1}{2}+
    \frac{\hat{p}^2_2}{2}+\frac{\hat{q}^2_2}{2}.
\end{equation}
The position and momentum operators in the Heisenberg representation
have the form~\cite{chernega2023dynamics}
\begin{eqnarray}
    \hat{q}^H_1(t) =\hat{q}_1\,\cos t +  \hat{p}_1\sin t, \quad
    \hat{q}^H_2(t) =\hat{q}_2\,\cos t + \hat{p}_2\,\sin t  \nonumber\\
    \hat{p}^H_1(t) =-\hat{q}_1\,\sin t + \hat{p}_1\,\cos t, \quad
    \hat{p}^H_2(t) =-\hat{q}_2\,\sin t + \hat{p}_2\,\cos t .
\end{eqnarray}
The center-of-mass tomogram can be rewritten as
\begin{equation}
    w(X|\mu_1,\mu_2,\nu_1,\nu_2;t) = w(X|\mu^H_1(t),\mu^H_2(t),\nu^H_1(t),\nu_2^H(t),t=0),
\end{equation}
where the time-dependent parameters are~\cite{chernega2023dynamics}
\begin{eqnarray}
    \mu^H_1(t) = \mu_1\,\cos t - \nu_1\,\sin t,\quad
    \mu^H_2(t) = \mu_2\,\cos t - \nu_2\,\sin t, \nonumber\\
     \nu^H_1(t) = \mu_1\,\sin t + \nu_1\,\cos t \quad
    \nu^H_2(t) = \mu_2\,\sin t + \nu_2\,\cos t.
\end{eqnarray}
In this way, the evolution of the center-of-mass tomogram for quadratic
Hamiltonians can be obtained by the corresponding time-dependent
transformation of the parameters of the initial tomogram. If the initial state
of the system is the entangled state~(\ref{eq:psi_ent}) then, after the
evolution, the state is described by
\begin{eqnarray}
    w_{\text{ent}}(X|\mu_1,\mu_2,\nu_1,\nu_2; t) =
\frac{e^{-X^2/\sigma}}{\sqrt{\pi\sigma}}\nonumber\\
   \times \left(\frac{1}{2} -\frac{\mu^H_1(t)\mu^H_2(t)+\nu^H_1(t)\nu^H_2(t)}{\sigma}+\frac{X^2}{\sigma}
    \left(1+\frac{2(\mu^H_1(t)\mu^H_2(t)+\nu^H_1(t)\nu^H_2(t))}{\sigma}\right)
    \right).
    \label{eq:ent_evolution}
\end{eqnarray}

The other example is a two-dimensional inverted oscillator with the
Hamiltonian
\begin{equation}
    \hat{H} = \frac{\hat{p}^2_1}{2}-\frac{\hat{q}^2_1}{2}+
    \frac{\hat{p}^2_2}{2}-\frac{\hat{q}^2_2}{2}.
\end{equation}
The center-of-mass tomogram of this system, which is initially in the
entangled state, has the form~(\ref{eq:ent_evolution}), where the parameters
are given by~\cite{chernega2023dynamics}
\begin{eqnarray}
    \mu^H_1(t) = \mu_1\,\cosh t  + \nu_1\,\sinh t , \quad
    \mu^H_2(t) = \mu_2\,\cosh t + \nu_2\,\cosh t;  \nonumber\\
     \nu^H_1(t) = \mu_1\,\sinh t + \nu_1\,\cosh t , \quad
    \nu^H_2(t) = \mu_2\,\sinh t  + \nu_2\,\cosh t.
\end{eqnarray}

\section{Cluster Tomography}
States of quantum systems with several degrees of freedom can be described by
cluster tomograms. The cluster tomogram for a system with three degrees of
freedom is defined as follows~\cite{dudinets2018center}:
\begin{eqnarray}
     w^{\text{cl}}(X,X_3|\mu_1,\mu_2,\mu_3,\nu_1,\nu_2,\nu_3)=
 \mbox{Tr} \left(\hat{\rho}\,\delta\left(X-\mu_1 \hat{q}_1-\nu_1 \hat{p}_1
 -\mu_2 \hat{q}_2-\nu_2 \hat{p}_2\right)\right.
 \nonumber\\\times
 \left.
 \delta\left(X_3-\mu_3 \hat{q}_3-\nu_3 \hat{p}_3\right)\right).
    \label{eq:cluster}
\end{eqnarray}
It is a conditional probability distribution of variables $X$ and $X_3$
related to the center-of-mass positions of the first and second subsystems
with two and one degrees of freedom, respectively. The positions are measured
in rotated and scaled reference frames determined by parameters $\mu_1$,
$\nu_1$, $\mu_2$, $\nu_2$, $\mu_3$, and $\nu_3$. The cluster tomogram for a
pure state with the wave function $\psi$ reads
\begin{eqnarray}
    w^{\text{cl}}(X,X_3|\mu_1,\mu_2,\mu_3,\nu_1,\nu_2,\nu_3) =
    \int dY_1dY_2\frac{\delta(X-Y_1-Y_2)}{4\pi^2|\nu_1\nu_2|}\times\nonumber\\
    \left|\int dq_1\, dq_2\,dq_3\psi(q_1,q_2,q_3)\exp
    \left(
    \frac{i\mu_1}{2\nu_1}q^2_1+\frac{i\mu_2}{2\nu_2}q^2_2+\frac{i\mu_3}{2\nu_3}q^2_3
    -\frac{iY_1}{\nu_1}q_1-\frac{iY_2}{\nu_2}q_2-\frac{iX_3}{\nu_3}q_3
    \right)
    \right|^2.
    \label{eq:cluster_pure}
\end{eqnarray}
Now we calculate the cluster tomogram of the $W$ state, which is an entangled
state of a three-dimensional oscillator of the form
\begin{eqnarray}
     \psi_{\text{W}}(q_1,q_2,q_3) = \frac{1}{\sqrt{3}}\left(\psi_0(q_1)\psi_0(q_2)\psi_1(q_3)+\psi_0(q_1)\psi_1(q_2)\psi_0(q_3)\right.
     \nonumber\\
\left.+\psi_1(q_1)\psi_0(q_2)\psi_0(q_3)\right).
\end{eqnarray}
The cluster tomogram of the state $W$ reads

\begin{eqnarray}
 w^{\text{cl}}_{\text{W}}(X,X_3|\mu_1,\mu_2,\mu_3,\nu_1,\nu_2,\nu_3)
 =\frac{2}{3\pi}\frac{e^{-\frac{X^2_3}{\sigma_3}-\frac{X^2}{\sigma_{12}}}}{\sqrt{\sigma_3\,\sigma_{12}}}\nonumber
 \left(
\frac{X_3^2}{\sigma_{3}}
+X^2\frac{(\mu_1+\mu_2)^2+(\nu_1+\nu_2)^2}{\sigma^2_{12}}\right.
 \\
\left.
+2\,X\,X_3\,\frac{\mu_1\mu_3+\nu_1\nu_3+\mu_2\mu_3+\nu_2\nu_3}{\sigma_3\,\sigma_{12}}+
\frac{(\mu_1-\mu_2)^2+(\nu_1-\nu_2)^2}{2\,\sigma_{12}}
\right),
    \label{eq:cluster_W}
\end{eqnarray}
where
$\sigma_3=\mu^2_3+\nu^2_3$ and $\sigma_{12}=\mu^2_1+\nu^2_1+\mu^2_2+\nu^2_2$.
The tomogram of the state W satisfies the 
normalization condition 
\begin{equation}
\int w^{\text{cl}}_{\text{W}}(X,X_3|\mu_1,\mu_2,\mu_3,\nu_1,\nu_2,\nu_3)\,dX\,dX_3=1.
\end{equation}
\section{Conclusions}

To conclude, we point out the main results of our work.

We considered the tomographic picture of quantum mechanics, where the states
of quantum systems are described by tomograms. The tomograms are fair
probability distribution functions in contrast to quasi-probability functions
like the Wigner function. This fact allows one to transfer some properties of
quantum systems to classical probability theory. In this paper, we focused on
the center-of-mass tomographic probability
distribution~\cite{arkhipov2005quantum}. We introduced entangled probability
distributions for the center-of-mass tomography; they correspond to entangled
states of quantum systems. Such probability distributions have not been
discussed in classical probability theory. We considered examples of
two-dimensional usual harmonic and inverted oscillators. Also, we studied
symplectic and cluster tomographic probability
distributions~\cite{dudinets2018center} for the oscillator states. We used the
method of integrals of motion~\cite{man2023quantum} to determine the time
evolution of tomograms.

We constructed new kinds of probability distribution functions, which earlier
have not been known in the probability theory describing the classical
randomness phenomena. One of such new probability distributions is the
center-of-mass probability distribution introduced for the description of
states of quantum harmonic oscillators. The other new probability distribution
is the cluster tomographic probability distribution. The introduced
probability distribution functions have a specific property. They describe
systems with many subsystems; for the center-of-mass tomography, only one
variable is enough but, for the cluster tomography, few random variables are
needed. Also, these probability distributions are able to describe the quantum
phenomenon of entanglement, which is not available for classical systems. In
view of these circumstances, we call these probability distributions the
entangled probability distributions. Such entangled probability distributions
have not been known in classical probability theory.

The approach to find and study the entangled probability distributions is
based on the possibility to construct probability representations of quantum
states, where the state wave functions have the properties that the linear
superpositions of the wave functions belong to a set of wave functions
determining the probability distributions. In classical mechanics, the
superpositions of classical trajectories do not have such a property. The
superposition of two solutions of Newton equation is not the solution to the
Newton equation. Thus, only in quantum world, the possibility appears to
obtain the entangled probability distributions describing the existing system
states, which are described by superpositions of the system wave functions. In
the future publication, we consider the possibility to construct classical
analogs of quantum center-of-mass probability distributions and study their
properties. Also, we will study the entropic characteristics of  the
center-of-mass probability distributions introduced for both quantum and
classical systems.

\vspace{6pt}




\section*{Funding}
This research received no external funding.




\section*{Appendix}

 Hereafter, we consider a system, which consists of two subsystems with the number degrees of freedom $N_1$ and $N_2$.
 In order to find the relation 
between tomograms of the system and its subsystems, we consider the
density operator of a separable state,
\begin{equation}
    \hat{\rho}_{12} = \sum_k p_k \,\hat{\rho}^{(k)}_1\otimes\hat{\rho}^{(k)}_2,
    \label{eqA:sep}
\end{equation}
where $p_k\geq 0$ and $\sum_k p_k = 1$.
The corresponding symplectic tomogram is given by the convex sum of the symplectic tomograms of the subsystems; it has the form
\begin{equation}
    w^{s}(\vec{X}_1,\vec{X}_2|\vec{\mu}_1,\vec{\mu}_2,\vec{\nu}_1,\vec{\nu}_2) = \sum_k p_k \, w^{s}_{1,\,k}(\vec{X}_1|\vec{\mu}_1,\vec{\nu}_1)\, w^{s}_{2,\,k}(\vec{X}_2|\vec{\mu}_2,\vec{\nu}_2).
    \label{eqA:symplectic}
\end{equation}
Here the variables are vectors with components $\vec{X}_1= \left(X_{1j}\right)$, $\vec{\mu}_1= \left(\mu_{1j}\right)$, $\vec{\nu}_1= \left(\nu_{1j}\right)$, where $j=1,2,...,N_1$ and $\vec{X}_2= \left(X_{2p}\right)$, $\vec{\mu}_2= \left(\mu_{2p}\right)$, $\vec{\nu}_1= \left(\nu_{2p}\right)$, where $p=1,2,...,N_2$.

The connection between the symplectic and the center-of-mass tomograms presented in~\cite{dudinets2018center} allows to obtain the relation between the center-of-mass tomogram of the system and its subsystems. Indeed, the center-of-mass tomogram of the state~(\ref{eqA:sep}) reads
\begin{equation}
    w(X|\vec{\mu}_1,\vec{\mu}_2,\vec{\nu}_1,\vec{\nu}_2) = 
    \int w^{s}(\vec{X}_1,\vec{X}_2|\vec{\mu}_1,\vec{\mu}_2,\vec{\nu}_1,\vec{\nu}_2)\,\delta\left(X-\sum_{j} X_{1j}-\sum_{p} X_{2p}\right)\,d\vec{X_1}\,d\vec{X_2},
\end{equation}
where the integral is taken over the components of the vectors $\vec{X}_1$ and $\vec{X}_2$. Next, we use Eq.~(\ref{eqA:symplectic}) to obtain
\begin{eqnarray}
    w(X|\vec{\mu}_1,\vec{\mu}_2,\vec{\nu}_1,\vec{\nu}_2) = \sum_k p_k\,
    \int \,\delta\left(X-\sum_{j} X_{1j}-\sum_{p} X_{2p}\right)\nonumber\\
    \times w^{s}_{1,\,k}(\vec{X}_1|\vec{\mu}_1,\vec{\nu}_1)\, w^{s}_{2,\,k}(\vec{X}_2|\vec{\mu}_2,\vec{\nu}_2)
    \,d\vec{X_1}\,d\vec{X_2}.
\end{eqnarray}
We express the symplectic tomograms of the susbsystems in terms of the center-of-mass tomograms, that is 

\begin{eqnarray}
    w^{s}_{1,\,k}(\vec{X}_1|\vec{\mu}_1,\vec{\nu}_1) = \frac{1}{(2\pi)^{N_1}} \int w_{1,\,k}(\zeta_1|\vec{k}_1\circ\vec{\mu}_1,\vec{k}_1\circ\vec{\nu}_1)\,e^{i\left(
    \zeta_1-\vec{k}_1\,\vec{X}_1
    \right)}\,d\vec{k}_1\,d\zeta_1,\\
    w^{s}_{2,\,k}(\vec{X}_2|\vec{\mu}_2,\vec{\nu}_2) = \frac{1}{(2\pi)^{N_2}} \int w_{2,\,k}(\zeta_2|\vec{k}_2\circ\vec{\mu}_2,\vec{k}_2\circ\vec{\nu}_2)\,e^{i\left(
    \zeta_2-\vec{k}_2\,\vec{X}_2
    \right)}\,d\vec{k}_2\,d\zeta_2.
\end{eqnarray}
Here $\vec{a}\circ \vec{b}$ denotes the vector with components $\vec{a}\circ \vec{b}=(a_j\,b_j)$, where $\vec{a} = (a_j)$ and $\vec{b} = (b_j)$. 

\begin{eqnarray}
w(X|\vec{\mu}_1,\vec{\mu}_2,\vec{\nu}_1,\vec{\nu}_2) =  \frac{1}{(2\pi)^{N}}\,\sum_k p_k\,
 \,\int
  w_{1,\,k}(\zeta_1|\vec{k}_1\circ\vec{\mu}_1,\vec{k}_1\circ\vec{\nu}_1)\,
  w_{2,\,k}(\zeta_2|\vec{k}_2\circ\vec{\mu}_2,\vec{k}_2\circ\vec{\nu}_2)\,
    \nonumber\\ \times
    \delta\left(X-\sum_{j} X_{1j}-\sum_{p} X_{2p}\right)\,
    e^{i\left(
    \zeta_1+\zeta_2-\vec{k}_1\,\vec{X}_1
    -\vec{k}_2\,\vec{X}_2\right)}
    \,d\vec{X_1}\,d\vec{X_2}
    \,d\vec{k}_1\,d\zeta_1
    \,d\vec{k}_2\,d\zeta_2,
    \label{eq:grand}
\end{eqnarray}
where $N = N_1+N_2$ is the number of degrees of freedom of the whole system.

We conclude that the center-of-mass tomogram of a separable state is presented
in the form~(\ref{eq:grand}).

\end{document}